\documentclass[useAMS,usenatbib]{mn2e}
\usepackage{mathtools}
\usepackage{graphicx}
\usepackage{color}
\usepackage{amsmath}
\usepackage{amssymb}

\def\h2{H$_2$}

\usepackage[breaklinks,colorlinks,
  urlcolor=blue,citecolor=blue,linkcolor=blue]{hyperref}

\def\Msun{\, M_{\odot}}

\def\MH2c{\, M_{\rm H_2,cell}}
\def\Msunpc2{\,\rm M_{\odot}\,pc^{-2}}

\title[Feedback and galactic environment in molecular clouds]{The roles of stellar feedback and galactic environment in star forming molecular clouds}

\author[Ramon Rey-Raposo et al.]{Ramon Rey-Raposo$^{1,2}$\thanks{E-mail:
rrr@astro.ex.ac.uk;}, Clare Dobbs$^{1}$, Oscar Agertz$^{2}$, Christian Alig$^{3}$\\
$^{1}$School of Physics \& Astronomy, University of Exeter, Stocker Road, Exeter EX4 4QL\\
$^{2}$Faculty of Engineering and Physical Sciences, University of Surrey, Guildford, GU2 7XH\\
$^{3}$ Universit\"ats-Sternwarte M\"unchen, Scheinerstr.1, D-81679 M\"unchen, Germany \\}
\begin{document}

\date{June 2016}

\pagerange{\pageref{firstpage}--\pageref{lastpage}} \pubyear{2016}

\maketitle

\label{firstpage}

\begin{abstract}
Feedback from massive stars is thought to play an important role in the evolution of molecular clouds. In this work we analyse the effects of stellar winds and supernovae (SNe) in the evolution of two massive ($\sim 10^6\,M_\odot$) giant molecular clouds (GMCs): one gravitationally bound collapsing cloud and one unbound cloud undergoing disruption by galactic shear. These two clouds have been extracted from a large scale galaxy model and are re-simulated at a spatial resolution of $\sim 0.01$ pc, including feedback from winds, SNe, and the combined effect of both. We find that stellar winds stop accretion of gas onto sink particles, and can also trigger star formation in the shells formed by the winds, although the overall effect is to reduce the global star formation rate of both clouds. Furthermore, we observe that winds tend to escape through the corridors of diffuse gas. The effect of SNe is not so prominent and the star formation rate is similar to models neglecting stellar feedback. We find that most of the energy injected by the SNe is radiated away, but overdense areas are created by multiple and concurrent SN events especially in the most virialised cloud. Our results suggest that the impact of stellar feedback is sensitive to the morphology of star forming clouds, which is set by large scale galactic flows, being of greater importance in clouds undergoing gravitational collapse.
\end{abstract}

\begin{keywords}
 galaxies: star formation -- ISM: clouds  -- hydrodynamics -- turbulence. -- gravitation
\end{keywords}

\section{Introduction}
Feedback from massive stars is considered to be an important source for driving turbulent motions in the interstellar medium (ISM) \citep[][]{MacLow2004} and in regulating the stellar content of galaxies throughout cosmic history via galactic scale winds \citep[][]{Silk2012, Hopkins2014,Agertz2015}. Details of this regulation are still unclear, and the importance of stellar feedback in regulating the rate of star formation inside of giant molecular clouds (GMCs) is a debated topic \citep[see e.g. reviews by][]{Padoan2014,Dobbs2013}. As stellar feedback acts on AU scales, but has cosmological implications, it is a difficult task to model simply due to the dynamical range of the problem.

Stellar feedback involves a number of different processes throughout the evolution of predominantly massive stars ($\gtrsim 8 \Msun$), from proto-stellar outflows at the beginning of their lives, to ionizing radiation, stellar winds and, finally, supernova (SN) type II events at the end of their lifetimes. The disruptive effect of the stellar feedback is possibly responsible for reducing the star formation rate in star forming clouds \citep{Goodwin2006,Bate2009,Dobbs2011,Kim2012b,Walch2012,Dale2013c} to observed rates \citep{Zuckerman1974,Krumholz2007a}. However, it has also been observed that feedback may trigger star formation on parsec scales \citep{Gouliermis2012}. Explaining this apparent conflict is crucial in order to understand the formation of multiple generations of stars in a molecular cloud, as the first population of stars will affect the creation of subsequent ones.

The effect of feedback in molecular clouds has been studied in a galactic context using global galaxy simulations \citep[e.g.][]{Dobbs2008,Hopkins2011,Agertz2013}. The star forming clouds emerging from these models present a wide range of morphologies and virial states making them, in principle, suitable for modelling the effect of feedback from stars at parsec scales. However, both the spatial and temporal limitations of these simulations make features such as wind blown bubbles (WBB) and shells difficult to resolve. 

Very detailed simulations of feedback disrupting isolated molecular clouds at sub-pc scales have been carried out \citep[e.g.][]{Dale2008,Walch2012,Colin2013,Padoan2015,Haworth2015}. Simulations including different feedback processes within the same molecular clouds have been performed with mixed results. \citet{Rogers2013a} modelled a small patch of a GMC affected by winds and SNe. They find that the effect of stellar winds is more important than the SNe in terms of dispersing the natal star forming gas, as the SNe energy escapes through the chimneys excavated by the winds. This indicates that SNe may impact the galaxy only on larger scales. This notion is also supported by other studies such as \citet{Fierlinger2015}, \citet{Rosen2014} and \citet{Ibanez-Mejia2015}. On the other hand, other studies find that the combination of ionising radiation from massive stars and stellar winds \citep[e.g.][]{Ngoumou2014,Dale2014} do not significantly change the outcome compared to the effect of the ionising radiation alone. An important caveat all these models have is that they are typically simplified and do not reflect galactic processes of cloud formation, a point which we will address here.

In previous work \citep[][ from now on Paper 1]{rrr2015} we studied the effect of the galactic context when simulating clouds with different morphologies to study the star formation process. We found that the inherited galactic velocity field has a considerable influence on the production of stars in the different clouds, even though they have roughly the same gas mass. However, the clouds included in Paper 1 were simulated under isothermal conditions and with no stellar feedback, the latter meaning that the star formation rates were unrealistically high, whilst we could not examine the relationship between galactic morphology and stellar feedback. In this paper we present an attempt to study the interplay between stellar feedback and galactic environment in molecular clouds resolving physics at scales of $\sim$ 0.01 pc. This paper is structured as follows: In Section 2 we describe the details of our simulations. Then we explain the main results of our work in Section 3. We compare to previous results in Section 4 and finish with a summary and conclusions in Section 5.

\section{Details of the Simulations}
\begin{figure*} 
\centering
\includegraphics[width=170mm]{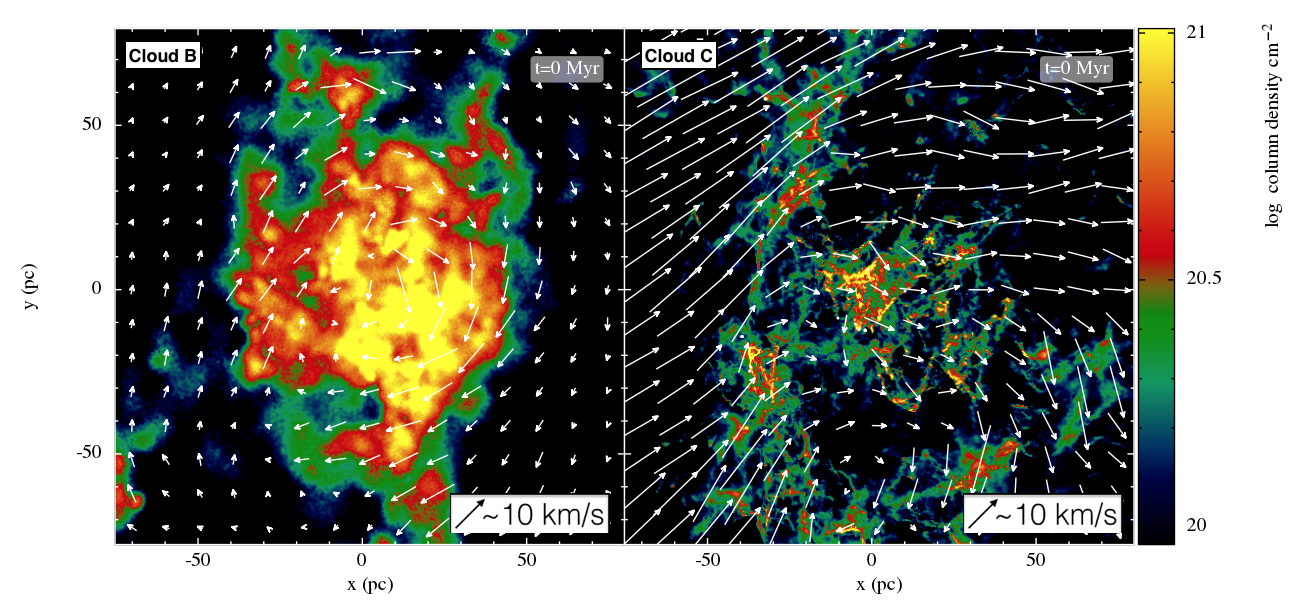}
\caption{Initial column density maps of Cloud B and Cloud C with the projected velocity field. Whereas in Cloud B the velocity field is dominated by the gravitational collapse, Cloud C experiences a strong shear inherited from the galactic simulation.} 
\label{fig:NF}
\end{figure*}
To study the relation between stellar feedback and the galactic environment, we focus on two clouds from Paper 1 (Cloud B and C) which respectively correspond to the most virialised, and the least virialised/most affected by galactic shear. The virial parameter is a dimensionless quantity that measures the boundedness of a mass of gas, by comparing the magnitudes of the kinetic and gravitational energy of the gas. It is often referred to as $\alpha$ \citep[e.g.][]{Dobbs2011a}:
\begin{equation}
\label{eq:VirialParameter}
\alpha  \equiv \frac{5R\sigma_v^2}{G M},
\end{equation}
where $\sigma_v$ is the gas velocity dispersion, $M$ the mass of the cloud, and $R$ its radius. A cloud is considered virialised if $\alpha \sim$ 1, and completely unbound if $\alpha > 2$. We extracted Cloud B from the galaxy simulation described in \citet{Dobbs2013a} after 250 Myrs of evolution. It includes self-gravity and a galactic potential with two components: a logarithmic disc potential following \citet{BinneyTremaine1987} and a spiral perturbation potential based on \citet{CoxGomez2002}. For the spiral potential they chose 2 arms, as that is the case for most symmetric galaxies. They included gas cooling and heating following \citet{Glover2007a,Glover2007b}, and H$_2$ chemistry \citep{Bergin2004}. Stellar feedback is injected using the same approach as in \citet{Dobbs2011b}. Cloud B has an approximate radius of 100 pc with a mass of 2.6 $\times 10^6$ M$_{\odot}$ and a virial parameter of $\alpha$ = 1.18. Cloud C is modelled from a spiral arm taken from a simulation by \citet{Dobbs2015}, that also includes gas cooling, H$_2$ chemistry and stellar feedback. Cloud C has similar size as Cloud B, with an approximate radius of 100 pc and a mass of 1.4 $\times 10^6$ M$_{\odot}$, but its virial parameter is substantially higher ($\alpha$ = 5.02). To extract the clouds we select the Lagrangian volume that encloses them in the galaxies. We obtain all the gas particles in that volume and increase the resolution of the gas following the method described in Paper 1. There are uncertainties in the feedback prescription of the galactic simulations, such as using a simple snowplough solution rather than incorporating all relevant feedback processes, the choice of IMF, and potentially overcooling (see \citet{Dobbs2015}). However, this model produces an ISM and molecular clouds with realistic properties compared to observations \citep{Dobbs2011b}\\

We re simulate these clouds using a modified version of the \textsc{sph} code \textsc{gadget2} \citep{Springel2005} that includes self gravity and sink particles \citep{Clark2008}. We include the chemistry of H$_2$ and CO following \citet{Pettitt2014}, which is based on \citet{Bergin2004} for the H$_2$ and \cite{Nelson1997} for the CO. We include the heating/cooling algorithm by \citet{Glover2007a}. We do not directly integrate the column densities (using for instance \textsc{Treecol} by \citet{TreeCOL}). Instead, we follow a simplified approach for both the chemistry and the cooling/heating using a fixed length of $L = 20$ pc to calculate column densities. This corresponds to the typical scales of the densest structures in the clouds. We use our reduced chemistry grid (H, H$_2$, CO, C$^{+}$) as parameters for the cooling subroutine updating its concentrations each timestep. We include subcycling (at least 200 times per timestep) to avoid truncation errors in the integration of the internal energy of each \textsc{sph} particle. We also prevent a change in the temperature larger than 30\% by limiting the timestep. Lastly, to avoid artificial fragmentation we set a temperature floor of $T_{\rm min}$ = 20 K \citep{Bate1997}.

\textsc{Gadget2} has known difficulties with resolving shear layers and turbulence. This problem is severe for sub-sonic turbulence, as shown in \citet{Agertz2007} and \citet{BauerSpringel2012}. However, density and velocity scaling present in supersonic turbulence, relevant in our simulated GMCs, are known to be of higher fidelity and compares well to grid based methods \citep{PHANTOM}, provided that comparable resolution is attained. We have also simulated the clouds in Paper1 without feedback using the \textsc{amr} code \textsc{ramses}, and see similar results to the \textsc{sph} simulations (see Rey-Raposo et al. in prep). 

We use 50 as the number of desired $\textsc{sph}$ neighbours (see e.g. \citet{Dehnen2012}), where each gas particle has a mass of $\sim 0.2\, M_\odot$ (depending on the cloud) and a gravitational softening of $\epsilon$ = 0.01 pc. The minimum \textsc{sph} smoothing kernel length is fixed so that $h_{\rm sml} \ge \epsilon$. Sink particles are formed by gas particles with convergent velocities ($\nabla \cdot \vec{v}_i < 0$), that are in a volume of radius $R_{\rm sink} = 0.1$ pc with a density higher than $\rho_{\rm sink} =1.6 \cdot 10^4$ cm$^{-3}$. Sink particles will accrete extra gas particles if the total momentum and energy of the sink and the gas particle corresponds to a gravitationally bound system \citep[for more details see][]{Jappsen2005}.

For both clouds we have run simulations with no feedback including the chemistry and ISM cooling/heating mentioned above. We consider feedback from winds and supernovae (SNe), testing the effect of the continuous action of winds in the clouds, and comparing it to the instantaneous injection of energy given by supernova explosions. We model both effects using the sink particles as sources. We do not know \textit{a priori} where these sinks will be created, as they form self-consistently in our clouds. Unlike the initial conditions based on uniform spheres, clouds extracted from galactic simulations present very complex structures. At sub-pc resolution this complexity makes the modelling of feedback very computationally demanding, as the number of sinks and therefore the number of feedback sources is high, greatly restricting the simulation time.

With our resolution, the sink particles represent star forming clumps, although we may refer to them as stars, sinks or sink particles indistinctly in this paper. We can not resolve individual stars, so we differentiate between the mass in sinks ($M_{\rm s}$) and the stellar mass contained in the sink ($M_{\rm *}$). We introduce two parameters $W_{\rm eff}$ and $SN_{\rm eff}$, where $W_{\rm eff}$ sets the fraction of O/B stars emitting winds and $SN_{\rm eff}$ the fraction exploding as SNe (see Appendix for more details). We chose a relatively high value for the parameter ($W_{\rm eff}$ = 0.75), whereby the effects of feedback are clearly visible on any cloud. We choose the same value of $W_{\rm eff}$ for the two clouds to be able to compare them, although the average mass of their sinks does differ. We select $SN_{\rm eff}$ = 0.50, considering that SN events happen in a longer time frame than the simulation, and if a sink particle hosts more than one massive star it is unlikely to see two SN events on the timescales of our simulations. We model the winds creating a \textsc{healpix} \citep{Gorski2005} sphere around each sink of radius 5 pc. We select a number of 48 \textsc{healpix} pixels, and we divide the momentum injection of winds \citep[following][]{Dale2008} by that number of pixels. To model SNe we use the same \textsc{healpix} scheme although within a larger radius (10 pc). We inject 10$^{51}$ erg into the gas particles inside the sphere, distributing it into 50\% thermal energy and 50\% kinetic energy. For more details about the feedback model, see Appendix \ref{appendixA}.

We use the final stage of the simulation with winds as the initial conditions for our SNe run. Hence, we stop the winds and we start injecting SNe, respecting the creation order of the sink particles until the end of the simulation. These relatively short timescales reflect the very long computational times of the simulations. We simulate the effect of winds in Cloud B for 1 Myr, and for Cloud C 2 Myr, to account for the different free-fall times of the clouds (3.5 and 7 Myr respectively). Lastly, to check the isolated effect of the SNe feedback we run two simulations with only feedback from SNe for clouds B and C, where the SNe explode 0.2 Myr after the creation of the sink. The summary of all the simulations is in Table \ref{tab:ch_4summary}. 

Apart from these simulations, we also run models of winds with $W_{\rm eff}$ ranging from 0.1 to 0.75, however we here present only the models adopting the highest value where the differences due to the feedback schemes, and the morphology of the clouds, are clearest. As our fiducial wind model is highly efficient, as discussed in Appendix A, we also explore the effect of a more conservative model. We here follow the parametrization of injected stellar wind momentum and energy from \citet{Agertz2013}, based on the stellar evolution code {\small STARBURST} 99 \citep{Leitherer1999} for single stellar populations (SSPs). We test this implementation in Cloud B for 1 Myr, assuming solar metallicity and the total conversion of wind thermal energy into kinetic energy.

In addition, we tested an algorithm that simulated the clouds by modelling the stellar mass following a more realistic IMF (based on \citet{Kroupa2001}). When a sink is created, we calculate its stellar mass (and time in the main sequence) by using the IMF as a random distribution. That stellar mass (and age) is fixed throughout the simulation and used to evaluate the feedback properties.  However, to observe a clear effect of winds of these simulations, we need to run them for a longer period of time ($\sim 5$ Myr), implying infeasible computing times. 
\begin{figure*} 
\centering
\includegraphics[width=185mm]{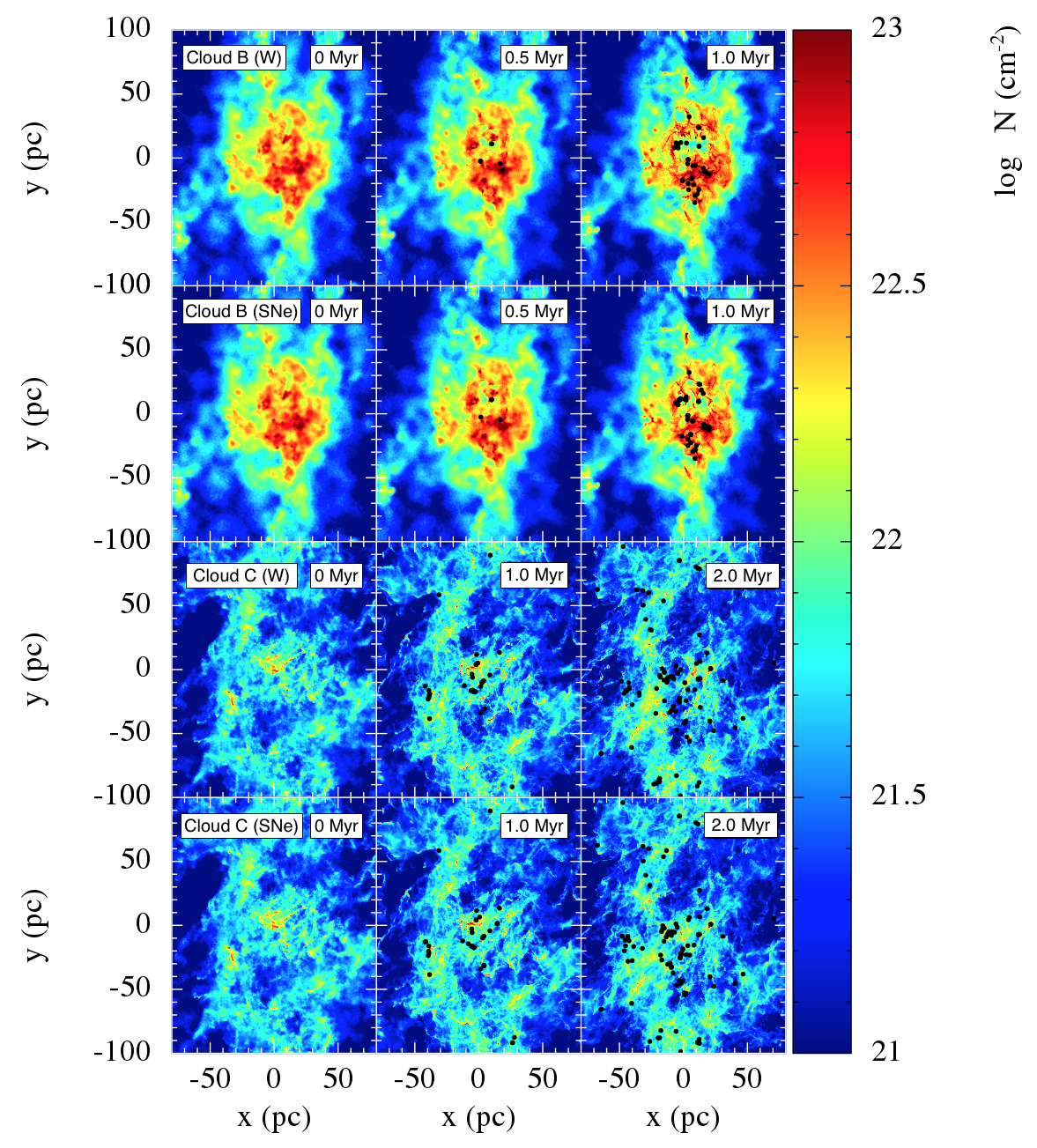}
\caption{Column density maps of the two clouds at different moments of the simulations. In the top row we show the evolution of Cloud B under the influence of winds. Winds (caused by the stars in the centre of the cloud) create a cavity. These WBB create a complex filamentary structure in the centre of the image. In the second row we observe the evolution of Cloud B subject to feedback from SNe. The feedback from SNe is not as effective disrupting the structure of the cloud as the winds. In the third row we show the evolution of Cloud C with feedback from winds. The effect of winds in this cloud dominated by galactic shear is less patent, although strong enough to create a small cavity visible in the last panel. Lastly, in the fourth row we present the temporal evolution of Cloud C affected by SNe. The effect of this kind of feedback is even less visible for this cloud.} 
\label{fig:t_evol}
\end{figure*}

\begin{figure*} 
\centering
\includegraphics[width=175mm]{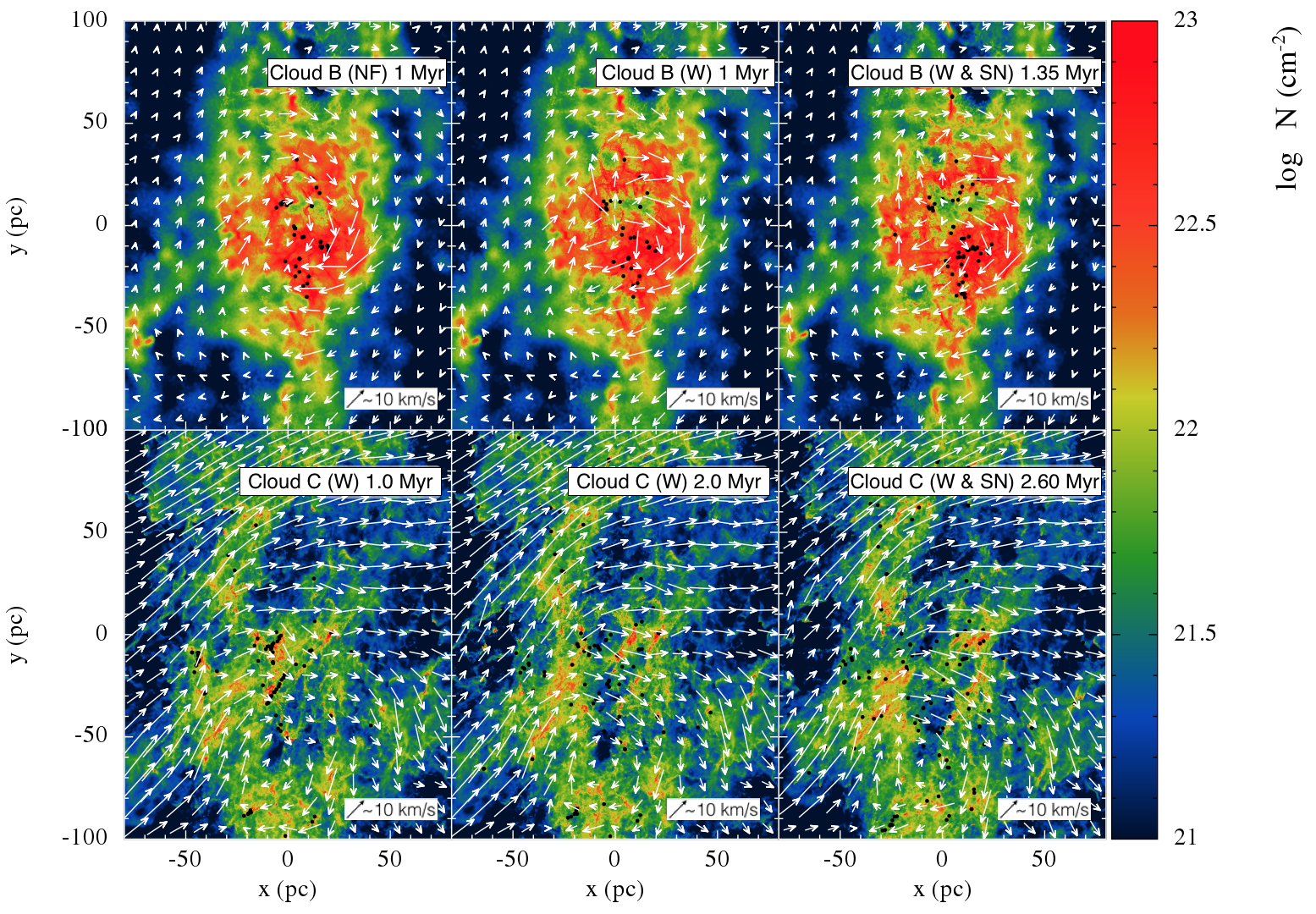}
\caption{Column density maps of the two clouds with velocity fields for the non feedback case (left panels), winds (central panels) and winds and SNe (right panels).}
\label{fig:vfield}
\end{figure*}
 
\begin{table}  
 
\caption{Summary of the simulations, including the mass, virial parameter, feedback mechanism and time when the simulation stops (t$_{end}$).}   
\centering 
 
\begin{tabular}{c c c c c c c c c} 

\hline\hline 

Cloud & Mass $(M_{\odot})$ &  $\alpha$ & Feedback & t$_{end}$ (Myr)\\ [0.5ex]

\hline 
Cloud B & $2.6 \times 10^{6}$ & 1.18 &  No feedback & 5.0\\

Cloud B & $2.6 \times 10^{6}$ & 1.18 &  Winds (D\&B08) & 1.0\\ 

Cloud B & $2.6 \times 10^{6}$ & 1.18 &  Winds (A+13) & 1.0\\ 

Cloud B & $2.6 \times 10^{6}$ & 1.18 &  SNe & 1.0\\

Cloud B & $2.6 \times 10^{6}$ & 1.18 &  SNe \& Winds & 1.35\\

Cloud C & $1.4 \times 10^{6}$ & 5.02 &  No feedback & 5.0\\

Cloud C & $1.4 \times 10^{6}$ & 5.02 &  Winds & 2.0\\

Cloud C & $1.4 \times 10^{6}$ & 5.02 &  SNe & 2.0\\

Cloud C & $1.4 \times 10^{6}$ & 5.02 &  SNe \& Winds & 2.6\\ [1ex] 

\hline 
 
\end{tabular} 
\label{tab:ch_4summary}
\end{table}

\section{Results}
In Fig. \ref{fig:NF} we show the initial projected velocity field for both clouds. As Cloud B is undergoing gravitational collapse, its velocity field is only noticeable in the very dense areas. There is a strong shear component in Cloud C that eventually disrupts the cloud and reduces its star formation rate ($SFR$). In the simulations without feedback, Cloud B creates a dense cluster of sinks in its centre, whereas in the absence of stellar feedback in Cloud C the production of sinks is not restricted to a particular area and they are formed in the filaments created by local collapse. 

In Fig. \ref{fig:t_evol} we show the temporal evolution of Cloud B (top rows) and Cloud C (bottom rows) for the simulations with feedback, presenting the clouds at different stages of their evolution. When winds are injected in Cloud B, in a short period of time (less than 1 Myr) a cavity with an approximate radius of 20 pc is created. This wind blown bubble (WBB) is formed mainly by the first sinks that can be observed in the top central panel of Fig. \ref{fig:t_evol}, and it is not observed in the non-feedback simulations. The rest of the sinks are formed at later stages and a great fraction of them are born in dense areas created by the winds in the periphery of the cavity, i.e. triggered star formation. The different shells created by sink particles are visible in the top right panel of the figure. These shells constitute an intricate network of filaments that in turn form new sinks in Cloud B. Nevertheless all the sinks in this cloud are restricted to the central region. The SNe have an observed impact on Cloud B, although their effect is less pronounced compared to that of the winds.
\begin{figure*} 
\centering
\includegraphics[width=175mm]{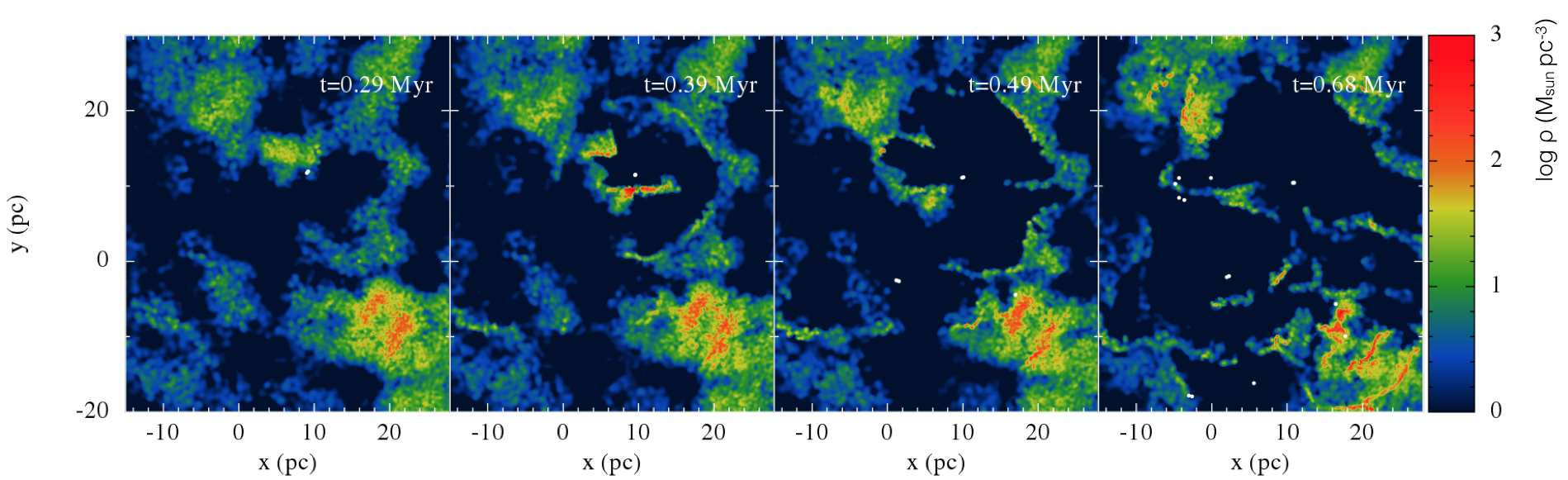}
\caption{Temporal evolution of a density cross-section of the wind blown bubble in the centre of Cloud B that is disrupted by winds over time.} 
\label{fig:puncture}
\end{figure*}

In the bottom panels of Fig. \ref{fig:t_evol} we present the evolution of the central area of Cloud C under the influence of feedback. The distribution of sinks in Cloud C is not restricted to the central region. Cloud C is affected by galactic shear over larger scales, and sink formation occurs in the filaments created throughout the entire cloud. The evolution of Cloud C seems to be largely unaffected by the stellar winds, at least during the first 1 Myr of the simulation. Even though the sinks are emitting winds from the moment they are created, their effect over the surrounding areas is less pronounced, and the inherited velocity field from the galaxy still dominates the dynamics of the cloud. The change in momentum caused by the winds represents approximately 10\% of the strength of the cloud's own velocity field ($\sim$1 km~s$^{-1}$ vs. $\sim$10 km~s$^{-1}$ for a typical gas particle), whereas in Cloud B the situation is the opposite; winds generate gas velocities that are an order of magnitude greater than the local velocity field. The winds weakly interact with the complex network of filaments which are stretched and deformed by the galactic shear. However, in the centre of the panel at 1.0 Myr we observe a shell of expanding gas surrounding a cluster of sinks that eventually creates a cavity in the lower centre of the cloud. The effect of the SNe is less clear than in Cloud B, but it is appreciable in the movie included in the additional material. 

\begin{figure*} 
\centering
\includegraphics[width=175mm]{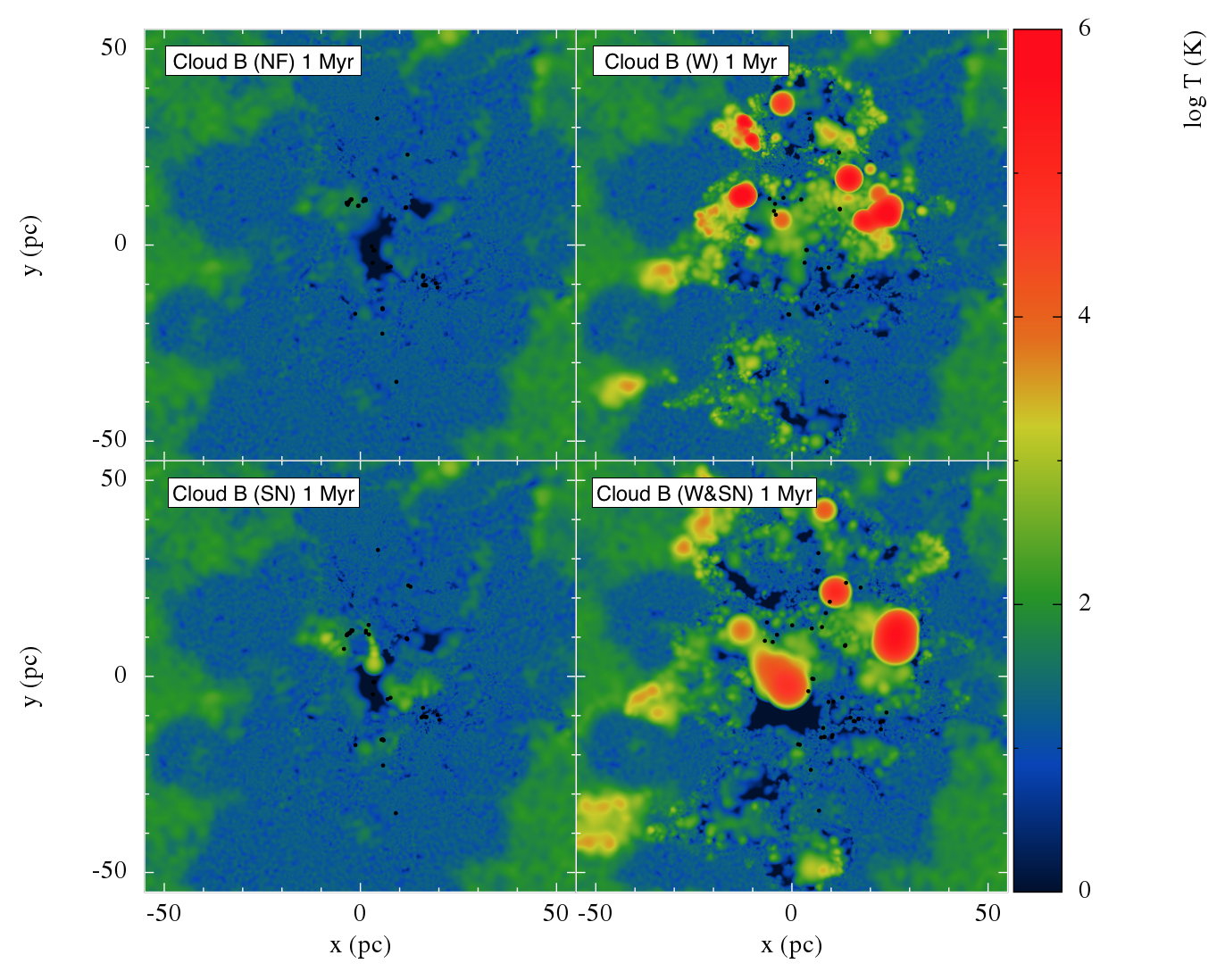}
\caption{Cross section of Cloud B (at $z = 0$) showing the temperature at 1 Myr for the non feedback run (top left) and the case with winds (top right). The simulations with SNe and combined feedback are shown in the bottom row. In the non feedback case Cloud B contains a cold central region surrounded by warmer material. Winds heat up the centre of the cloud to temperatures up to $T \sim 10^6$ K, whereas the increase in temperature caused by a SN is a localised event.}
\label{fig:2CloudsT}
\end{figure*}

\subsection{Effect of the Stellar Feedback on the Velocity Field of the Clouds}
In Fig. \ref{fig:vfield} we show the projected velocity field over the $z$-axis for Cloud B (top row) and Cloud C (bottom row) overlaid on the column density maps of these clouds. The left panels present the velocity field for the non feedback simulations, the central panels the velocity field at the end of the winds simulations and the right panels the velocity field for the runs with combined winds and SNe end. For Cloud B, there is a change in the velocity field in the central region, as the winds are dispersing the gas, reducing gravitational collapse. In the non feedback case the velocity field of the cloud is dominated by gravitational collapse. When SNe are introduced the instantaneous nature of the multiple explosions distorts this spherical pattern. 

The global velocity field of Cloud C does not vary substantially from the non feedback case, featuring a strong shear flow from the north-east to the south-west of the cloud (the values for the velocity in this cloud are an order of magnitude larger than for Cloud B). Furthermore the sinks in this cloud are less massive and therefore the wind strength is reduced (Eq. \ref{eq:Dale08}). Hence, the velocity field is practically unaffected by winds, and SNe also do not modify greatly the inherited galactic shear. In both clouds, the shocked gas from the winds punctures the main wind blown bubble and escapes through the diffuse areas of the clouds (see Fig. \ref{fig:puncture}). This is also observed in other works such as \citet{Rogers2013a}. 

\begin{figure*} 
\centering
\includegraphics[width=175mm]{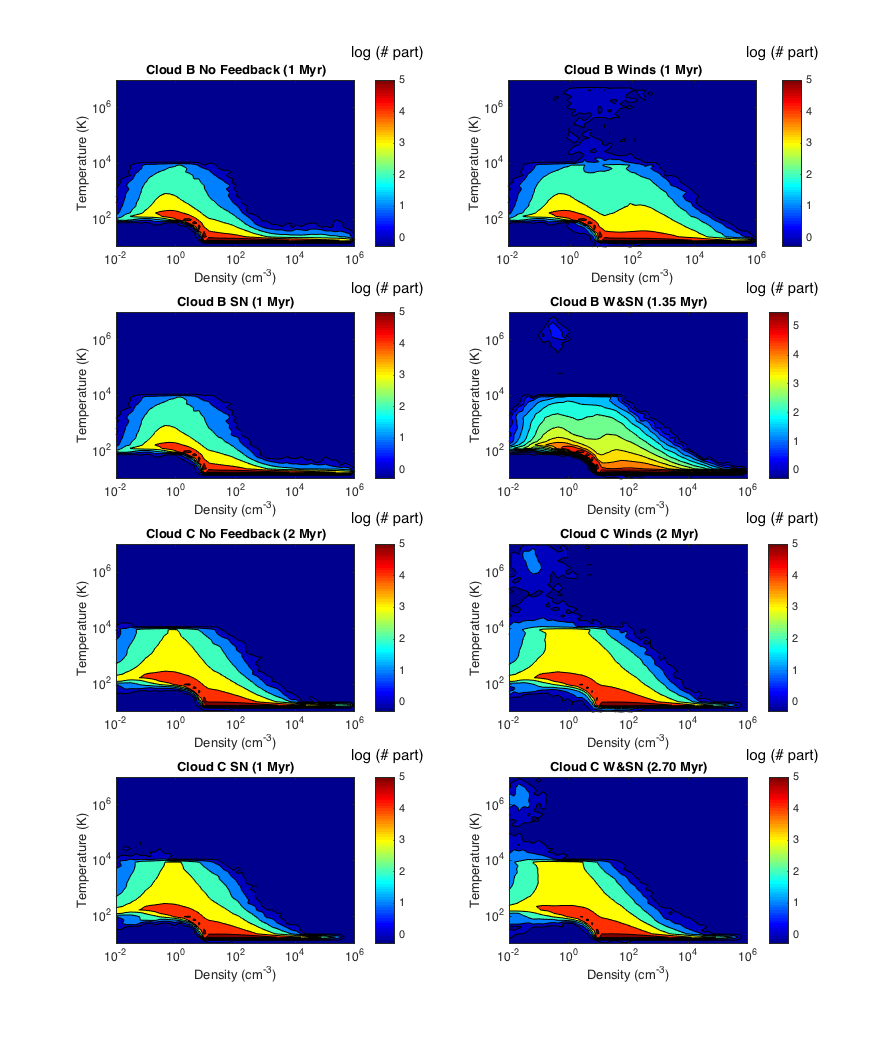}
\caption{Temperature vs. density histogram of Cloud B (at 1 Myr for the non feedback, winds and SNe, and at 1.35 Myr for the combined feedback) and of Cloud C (at 2 Myr for the non feedback, winds and SNe, and at 2.70 Myr for the combined feedback). In general for both clouds, the non feedback case most gas is at very low temperatures ($T \sim T_{\rm min}$ = 20 K), but there is a wide scatter reaching temperatures up to 10$^4$ K where cooling from HI becomes effective. In the winds run most gas is still at very low temperature ($T \sim T_{\rm min}$), but in this case there is a fraction of shocked hot gas at temperatures of the order of $T \sim 10^6$ K. The SNe heat up a small portion of the gas too insignificant to see in the histogram. For the combined run we still observe the remains of hot gas from the winds.} 
\label{fig:hist_Rho_T}
\end{figure*}

\subsection{Effect of the Stellar Feedback on the Thermal Properties of the Clouds}
In  Fig. \ref{fig:2CloudsT} we present a temperature cross section through the central plane of the galaxy ($z = 0$) of Cloud B at 1 Myr for the non feedback run, SNe and winds run. The combined feedback run is shown at 1.35 Myr. If feedback is not present the cloud is cold, especially in the central part where the temperature has values close to $T \sim 25$ K, corresponding to the areas of higher molecular abundance. This molecular central region is surrounded by gas at higher temperatures ($T \sim 200$ K). 

When feedback is introduced, winds shock the surrounding medium and the gas reaches temperatures up to $T \sim 10^6$ K. This increase in the central temperature of the cloud is not maintained in the simulations with SNe, where the effect is more locally restricted and we need to present a cross section at the exact depth where the SNe is happening at that particular moment. In the combined run there is still a substantial fraction of hot gas in the central region, the gas affected by the winds is able to cool down. The effects of several exploding SNe are now clearly visible. In the non feedback scenario, Cloud C is also substantially cold on average. However, the shear promotes the mixing of hot gas from the periphery of the cloud into the central parts. The overall effect of feedback is similar to Cloud B, where winds are found to heat up the regions surrounding the sinks.

We show temperature histograms of the clouds in Fig. \ref{fig:hist_Rho_T}. For the case without feedback the clouds present temperatures in the range of $T\sim 20 - 10^4$ K. For Cloud B the inclusion of feedback does not change the temperature distribution in that range. However there is a small amount of gas at temperatures higher than $10^6$K. As this gas cools, a stream extends downs to higher densities and lower temperatures. This cooling gas produces a second heap on the diagram having a significant population of relatively dense gas ($\sim 100$ cm$^{-3}$) at warm temperatures (over 1000 K). As shown in the cross section (Fig. \ref{fig:2CloudsT}), there is only a small amount of gas heated up by the SNe, which has practically no effect on the histogram. 

In the combined runs we still observe some gas particles at very high temperatures, but even though we have SNe feedback, the fraction of hot gas is reduced as winds are no longer active, leading to less shock heated gas in the centre of the cloud. The fraction of warm diffuse gas is higher for Cloud C. Winds also shock the gas for Cloud C promoting the particles to the upper part of the histogram as for Cloud B. The effect of the SNe is again less appreciable.

\subsection{Energy Budget of the Clouds}
\begin{figure*}
\centering
  \includegraphics[angle=0,width=170mm]{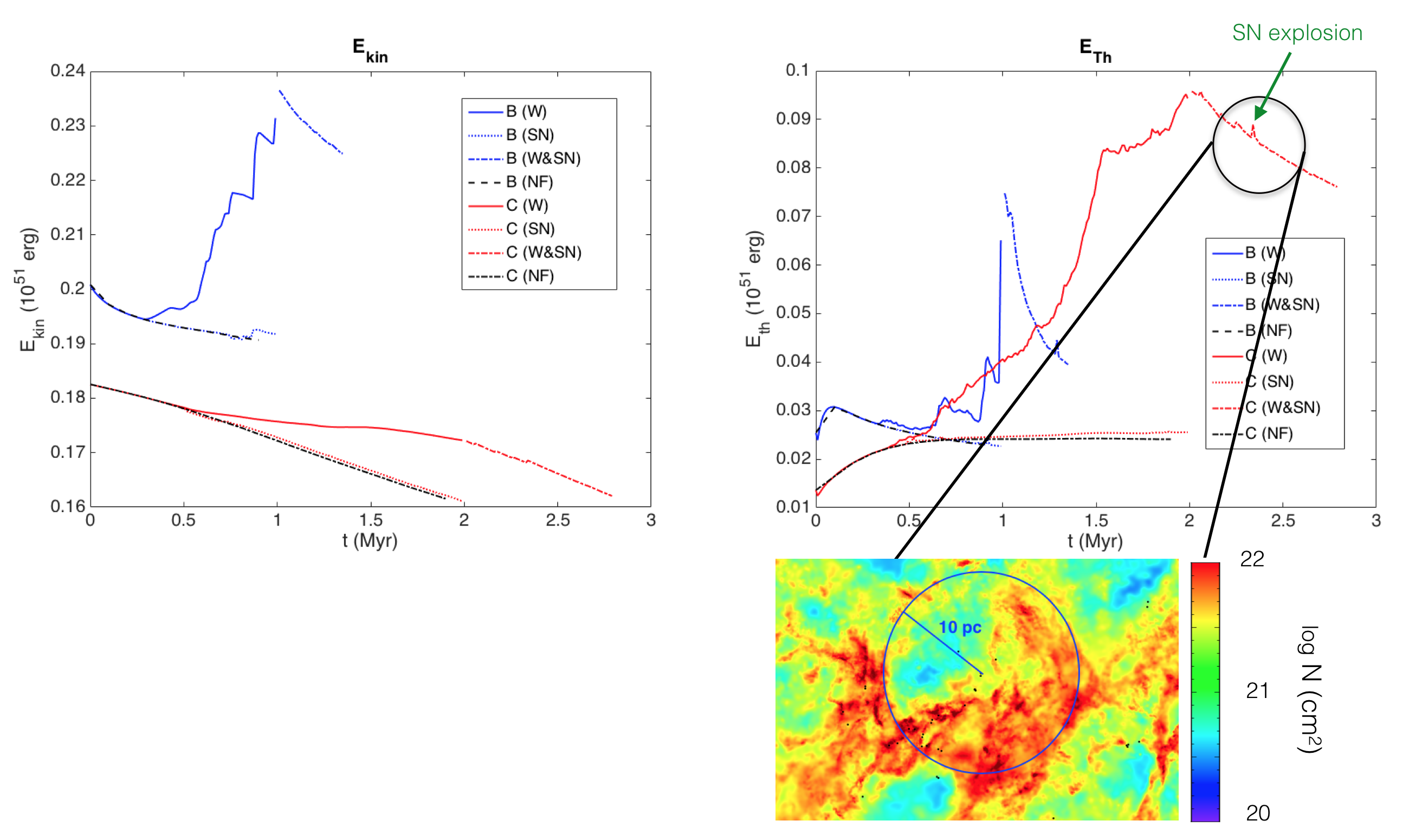}
  \caption{Total thermal (left panel) and kinetic (right panel) of the simulations. The injection of energy of the SNe is quickly dispersed via cooling. Only the winds cause an impact, increasing the energies of both clouds. We have selected the first SNe events after the winds in Cloud C (slightly visible), and calculate the thermal energy injection with extremely high time resolution. The thermal energy of the cloud is doubled although it quickly decreases as the gas cools down. We also include a column density plot of the area in which the SN feedback is injected. There is a difference in the density field around the sink that covers 2 orders of magnitude. }
  \label{fig:energy}
\end{figure*}
To examine why the effect of winds is much more pronounced than the SNe we present the total thermal and kinetic energy of the clouds in Fig. \ref{fig:energy}. We express the energies of the clouds in $E^O_{\rm SN} =  10^{51}$ erg units. Once the gas temperature has settled (we initialise the clouds at $T$ = 50 K), the thermal energy for the non feedback case remains relatively constant. Note that at 1 Myr both clouds have the same thermal energy, even though Cloud B is almost twice as massive as Cloud C. This is because Cloud C possess more gas at a higher temperature. For the kinetic energy there is a decrease due to the fact that more gas is captured in sinks, and is not accounted for. The kinetic energy of both clouds is an order of magnitude larger than the thermal. When integrating the total amount of energy for the clouds we find that for Cloud B the winds inject 5\% more kinetic and 13\% more thermal energy than the SNe. In Cloud C winds are more efficient than SNe injecting thermal energy (50\%). The increase in kinetic energy is only 3\% for this cloud.

The behaviour for the SNe runs (dotted lines) is again very similar to the non-feedback scenarios, suggesting that the energy (thermal and kinetic) injected is quickly radiated away. We have selected the first SNe events after the winds in Cloud C and recalculated the thermal energy of the cloud with a temporal precision of months. The thermal energy of the cloud is nearly doubled, although the energy is quickly radiated away. In the top right panel of Fig. \ref{fig:energy} we include the column density field of the area in which the feedback from the first SNe will be injected. The density varies in the affected area, at least, by two orders of magnitude. Hot shocked gas (T $\sim 10^6$ K) has different cooling times ranging from $\sim 100$ yr for $n \sim 10^3 $ cm$^{-3}$ (see e.g. \citet{Agertz2013,Sutherland1993}) to $\sim$ 1 yr for the very dense areas ($n \sim 10^6 $ cm$^{-3}$). The cooling in those dense regions is very efficient, and therefore the gas around the sink is able to lose the injected energy swiftly. 

Winds shock the ISM, increasing considerably the thermal energy of both clouds, as the continuous injection of energy impedes the cooling of the hot gas. For both clouds there is a considerable increase in the thermal energy of the cloud. However, whilst the kinetic energy of Cloud B is increased by the winds, the effect for Cloud C is not so strong. The shear dominated velocity field of Cloud C, helps dispersing the injected momentum in the cloud. This difference in the effect of feedback is in agreement with the difference in the star formation rate of the clouds (see section \ref{ssection:SFR}). As suggested in previous sections, the characteristics of the cloud affect the way feedback interacts with the gas, and how it couples with the ISM. The effect of winds is also clear in the kinetic energy of the clouds. Again for Cloud B this is greater than for Cloud C.

\subsection{Effect of the Stellar Feedback on the Virialisation and Molecular Fraction of the Clouds}
\begin{figure}
\centering
\includegraphics[angle=0,width=90mm]{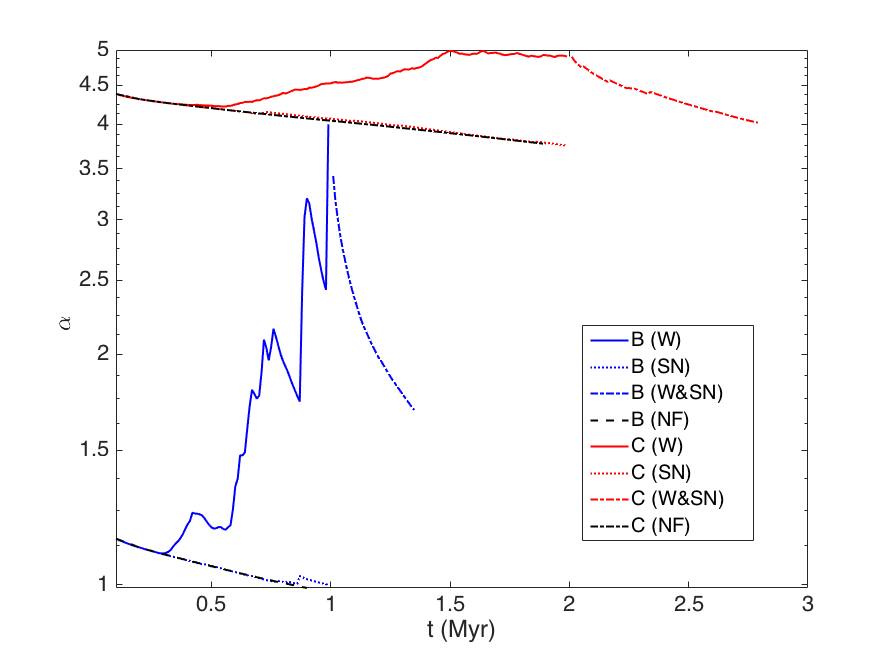}
\caption{Variation of the virial parameter with time for clouds B and C. The virial parameter decreases for the non-feedback case, as the clouds become more gravitationally bound. Winds eject more energy in the ISM increasing the velocity dispersion of the gas, and resulting in a growing virial parameter. The effect of SNe is negligible compared to the non-feedback scenario.}
\label{fig:snavst}
\end{figure}

In Fig. \ref{fig:snavst} we show the variation of the virial parameter over time. To calculate the virial parameter we calculate the velocity dispersion of all the gas particles at each timestep, keeping the other parameters in Eq. \ref{eq:VirialParameter} constant. We therefore neglect the change in size of the cloud over time but this effect is smaller than 10\% for the simulation times considered here. Cloud B starts as a highly virialised cloud, whereas Cloud C starts from an unbound state. In the non feedback scenarios, the clouds become more virialised as they collapse and form more sink particles reducing the velocity dispersion of the gas (as sinks do not contribute to the calculation) and therefore making the virial parameter smaller. 

Winds shock the gas around the sinks, increasing significantly the velocity dispersion of the clouds. Again the effect is stronger for Cloud B than for Cloud C. In the moment the first sink is created in Cloud B ($t \sim 0.3$ Myr) the cloud departs from its virialised state, becoming non-virialised for $t > 0.8$ Myr. For the gravitationally bound Cloud B ($\alpha \sim 1$), winds disrupt the cloud and increase $\alpha$ to a similar virial state to Cloud C ($\alpha \sim 5$). 

For Cloud C where winds are less effective, $\alpha$ does not increase more than 10-15\%. The effect of the SNe is smaller, seeming to have little impact on the virial state of the cloud. The de-virialisation effect of feedback has also been observed in other simulations like \citet{Colin2013}.
\begin{figure}
\centering
\includegraphics[angle=0,width=90mm]{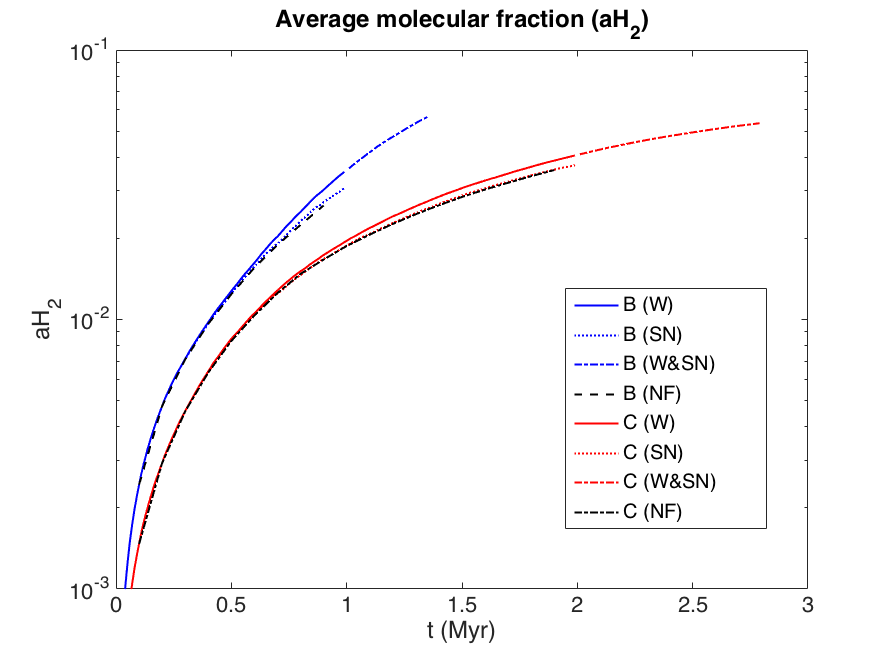}
\caption{Variation of the average molecular fraction with time for clouds B and C. The stellar feedbacks seems to have little impact on the molecular component of the clouds.}
\label{fig:snaH2t.eps}
\end{figure}

We also analyse the temporal evolution of the density averaged molecular factor aH$_2$ in Fig. \ref{fig:snaH2t.eps} for the two clouds. We define the molecular factor as
\begin{equation}
\label{eq:aH2}
aH_2 = \frac{n(H_2)}{n(H) + 2n(H_2)}, 
\end{equation}

where $aH_2$ varies between 0 and 0.5, and $n(H_2)$, $n(H)$ are the numerical abundances of H$_2$ and H respectively. We do not find great differences between the feedback runs for a given cloud, suggesting that feedback does not have a great impact on the chemistry over the timescales probed by our simulations. It is worth noting that Fig. \ref{fig:snaH2t.eps} shows the mass-averaged molecular fraction, and in very dense areas the cooling is highly effective and therefore gas quickly cools retaining a high molecular fraction. The cloud may become unbound ($\alpha >$ 2) and still retain its molecular state. 

\subsection{Effect of the Stellar Feedback on the Star Formation in the Clouds}
\label{ssection:SFR}
\begin{figure} 
\centering
\includegraphics[angle=0,width=90mm]{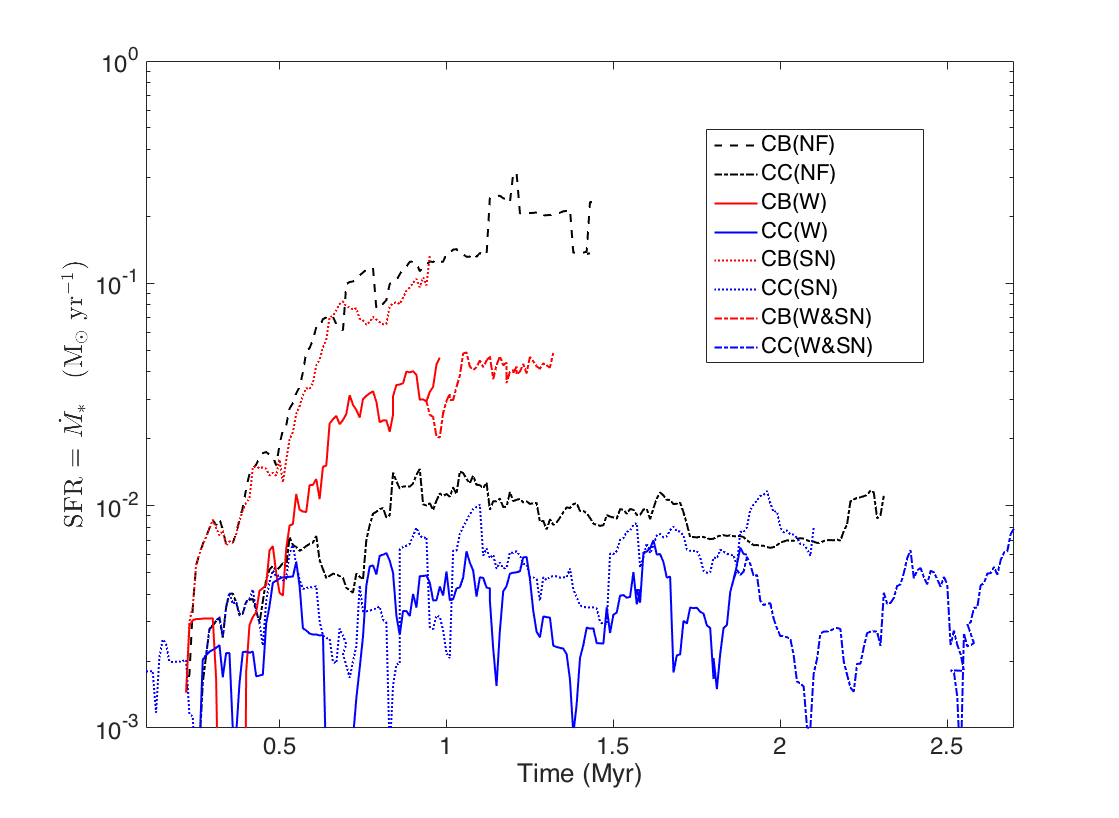}
\caption{Star Formation Rate for the 2 clouds. We show with dashed lines the non-feedback runs, in solid the simulations with winds and with a dotted line the cases with SNe. Winds reduce the $SFR$ in both clouds whereas the effect of SNe is not so clear.}
\label{fig:snSFR2.eps}
\end{figure}

\begin{figure}
\centering
\includegraphics[angle=0,width=90mm]{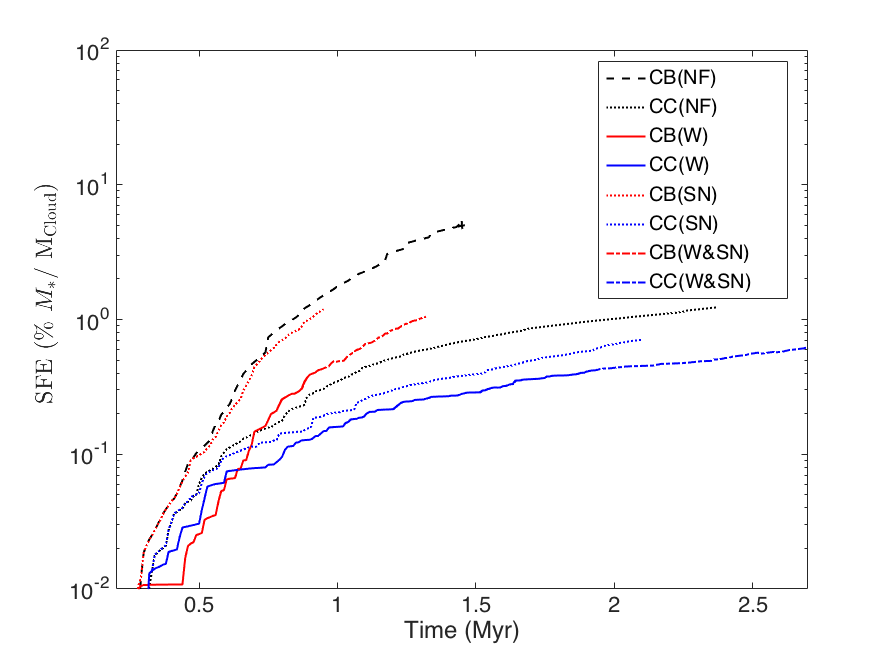}
\caption{Star Formation Efficiency for the 2 clouds. We show with dashed lines the non-feedback runs, in solid the simulations with winds and with a dotted line the cases with SNe.}
\label{fig:snSFE2.eps}
\end{figure}

We define the star formation rate ($SFR$) as $SFR(t) =  \dot{M}_{\rm s}(t)$, where $\dot{M}_{\rm s}(t)$ is the time derivative of the mass contained in sinks. We use bins of 0.01 Myr. In Fig. \ref{fig:snSFR2.eps} we show the $SFR$ for the simulations. We define the star formation efficiency as $SFE = M_{\rm s} / M_{\rm cloud}$ (where $M_{\rm cloud}$ is the initial masss of the cloud), and this is shown in Fig. \ref{fig:snSFE2.eps}.

In general, winds reduce the $SFR$, mainly by stopping accretion onto sinks (see also \citealt{Dale2008}), but also through dispersing the gas. SNe also reduce the SFR, although their effect is less powerful than the winds. The high concentration of SNe in the central area of Cloud B causes the collision of ejecta from different events, increasing the density and triggering star formation.

The effect of the winds is different depending on the cloud. For Cloud B the reduction in the $SFR$ is nearly 100\% while for Cloud C the reduction is around 50\%. In Cloud B winds act opposing the global gravitational collapse of Cloud B, whereas in Cloud C the galactic shear is still dominating. As we previously discussed, for Cloud B winds are $\sim 10$ times stronger than the inherent velocity field, whereas in Cloud C the inherited galactic velocity field is more powerful than the winds. When SNe are introduced the $SFR$ does not reach that of the non feedback runs. 

\begin{table}  
\caption{Star Formation Efficiencies for Simulations.}   
\centering 
\begin{tabular}{c c c c c} 

\hline\hline 

Cloud & Feedback & t$_{end}$ (Myr) & SFE \\ [0.5ex]

\hline 
Cloud B &  No feedback & 1.0 & 2\% \\

Cloud B &  Winds (D\&B 2008) & 1.0 & 0.5\% \\ 

Cloud B &  Winds (A 2013) & 1.0 & 1\% \\

Cloud B &  SNe & 1.0 & 1\% \\

Cloud C &  No feedback & 2.0 & 1\% \\

Cloud C &  Winds & 2.0 & 0.6\% \\

Cloud C &  SNe & 2.0 & 0.7\%  \\

\hline 
 
\end{tabular} 
\label{tab:ch_SFR}
\end{table}

The $SFR$ is relative to the initial mass of the cloud. To compare with observed clouds we use the star formation efficiency rate $SFER(t) = SFR(t)$ / M$_{\rm cloud}$ (as defined in \citet{Dale2014}). With winds, we find values of 0.005 Myr$^{-1}$ for Cloud B and 0.003 Myr$^{-1}$ for Cloud C. This is comparable to or even lower than observations, where $SFER$ $\sim$0.015 - 0.030 \citep{Evans2009}. In Fig. \ref{fig:snSFR2.eps} we present the $SFE$. The $SFEs$ for Cloud C ($\sim$0.01) and Cloud B ($\sim$0.02) are also smaller compared to other computational works (in \citet{Dale2014} they are of the order 0.05 - 0.20). But we note that our simulation times are relatively short, and that particularly for Cloud C (where in Paper 1 we saw that the star formation rate grew with time), the star formation rate may increase. Observations of well known HII regions like 30 Doradus, show that the star formation process occurs in different stages, each generation affecting the next one \citep{deMarchi+2011,Doran+2013}. In this region, an older generation of stars ($\sim$ 20 Myr) produces feedback which triggered the formation of the bright cluster in the centre of the region containing massive stars that are a few Myr old. These stars also coexist with protostellar objects covered in dusty envelopes that will eventually disperse liberating newly born stars within them. We observe the existence of different generations of stars in the models in Paper 1 that cover longer timescales ($\sim $ 10 Myr). However when we include feedback we are restricted by the short simulation times, and therefore the $SFE$ only accounts for the first stages of the star formation process.  

\subsection{Effect of the Stellar Feedback on the Mass distribution in Sink Particles}
\begin{figure*} 
\centering
\includegraphics[width=175mm]{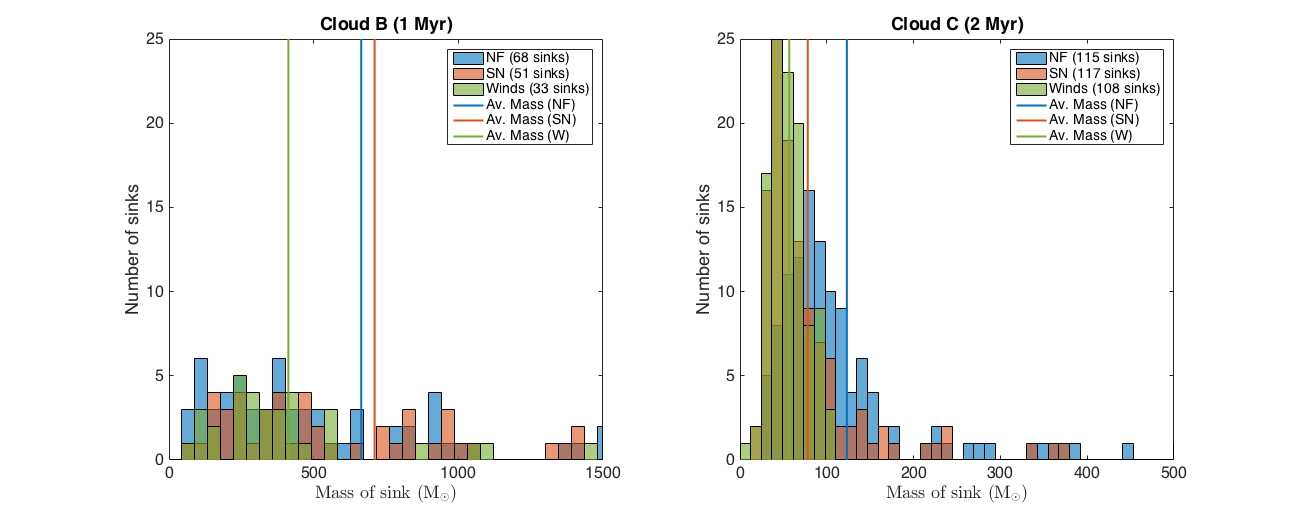}
\caption{Histogram of the mass distribution for the different simulations. The vertical lines represent the average mass for each distribution. The non-feedback run includes a high-mass tail produced by the continuous accretion of gas on the sinks. This tail is effectively suppressed by winds, but not supernovae. The winds reduce the mass of the sinks in both clouds, and also reduce the number of them in Cloud B. The winds in Cloud C do not significantly reduce number of sinks as the velocity field dominates and the winds have less effect on the cloud morphology.} 
\label{fig:sinks}
\end{figure*}

In Fig. \ref{fig:sinks} we show the histograms of the mass distribution of sinks for each cloud, containing the data from the 3 runs. We have also included a vertical line representing the average sink mass for each run. In the left panel of the figure we depict the distribution of sinks for Cloud B at 1 Myr (we do not consider the combined run with winds and SNe). The average sink mass for the non-feedback scenario is high ($\sim 800 M_{\odot}$), as the sink distribution has a well-defined high mass tail. 

The inclusion of winds severely reduces the mass of sinks, moving the average to $300 M_{\odot}$. This tail is the result of continuous accretion of gas by the sink particles, as this process is interrupted when the stars emit winds. This assumption is also supported by the presence of the high mass tail in the SNe run (where the accretion is not totally interrupted). In the right panel of Fig. \ref{fig:sinks} we present the same data for Cloud C at 2 Myr. The average mass of sinks of Cloud C is smaller ($\sim 120 M_{\odot}$) for the non-feedback scenario, with a long tail of high mass sinks. Again this tail is reduced especially with the inclusion of winds.

We also find differences between the clouds. Although feedback in general reduces the mass in sinks, in Cloud B the average sink mass is reduced by a factor of 3 when winds are introduced, for Cloud C, it is only reduced by a factor 2. The number of sinks remains practically constant for Cloud C ($\sim 100$), whereas for Cloud B the inclusion of feedback halves the number of sink particles. We emphasize that the simulation parameters for the sinks are the same for every run, therefore the differences between the clouds in their star forming histories are caused by the initial conditions. With no feedback Cloud B tends to quickly collapse and form large sinks that continue accreting gas. Only winds stop this accretion, but in both runs with feedback the creation of new overdense areas in the periphery of the bubbles causes the formation of new sink particles. Affected by galactic shear in Cloud C, less massive sinks are created in the non feedback case. The global velocity field dominates the evolution of the cloud (even with the inclusion of feedback) and sinks are created in the filamentary structures of the cloud driven by the inherited turbulence. 

\subsection{Effect of the Different Winds Prescription on the Clouds}
Our model uses strong winds (based on \citet{Dale2008}), as we assume that 75\% of the mass of a sink particle is generating feedback. Also the strongly non-linear dependence on the stellar mass in Eq. \ref{eq:Dale08} makes the model sensitive to correctly modeling the number and masses of high mass stars, something that our sink approach is not designed to do. We use an enhanced model for the SNe letting a SNe explode 0.2 Myr after the sink is created for the SNe run, and 0.7 Myr in the combined run.
  \begin{figure*}
  \centering
   \includegraphics[angle=0,width=170mm]{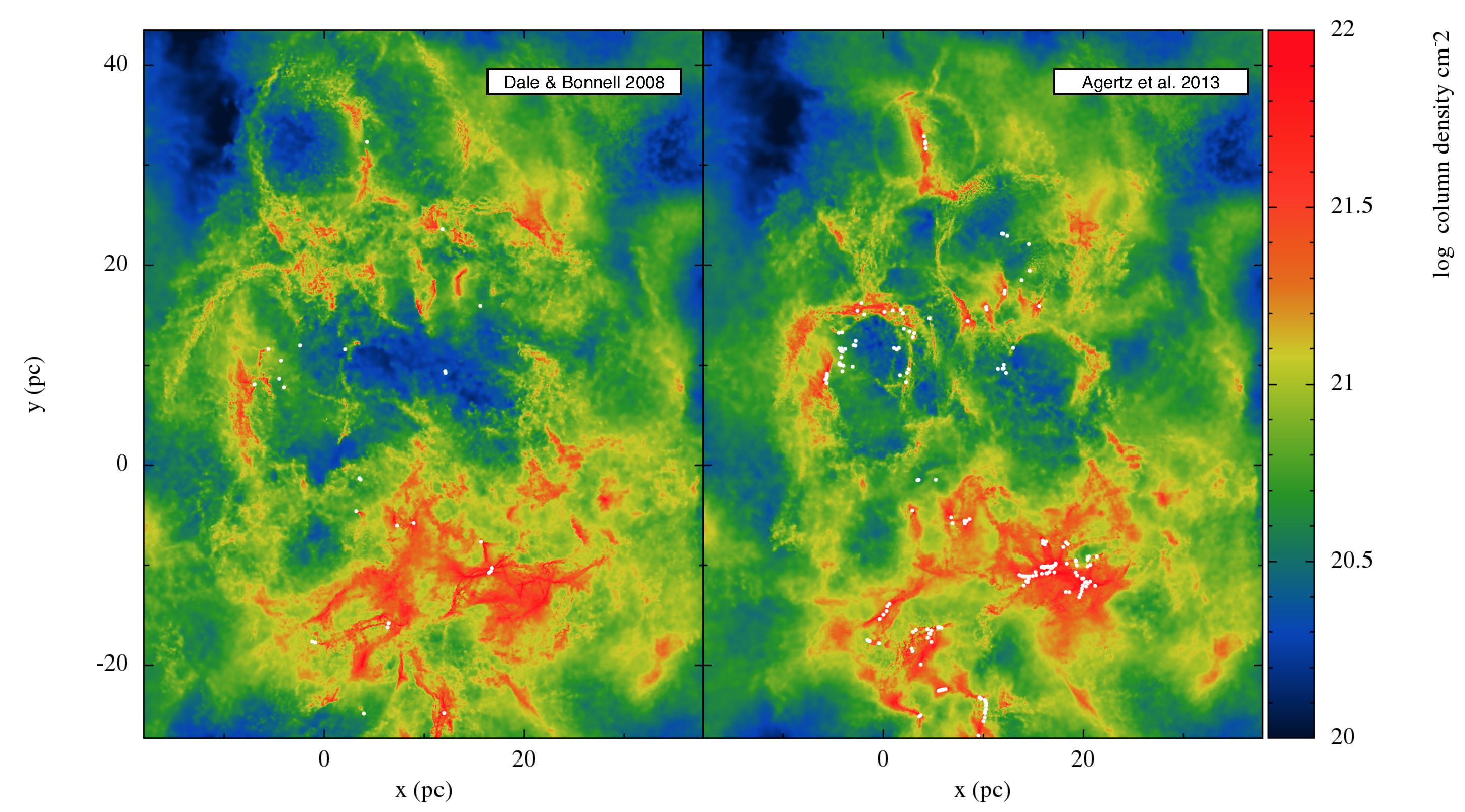}
    \caption{Column density map of Cloud B with winds modelled using \citet{Dale2008} (left panel), and using \citet{Agertz2013} (right panel) at the end of the simulation (1 Myr).}
    \label{fig:OA}
 \end{figure*}

To confirm that the effect of the winds in the clouds is more prominent than the SNe, we have run an adapted version of the algorithm from \citet{Agertz2013}. In this work they calculate the strength of the stellar winds from single stellar populations (SSPs) using data from {\small STARBURST99} \citep{Leitherer1999}, representing a more conservative approach to the total amount of momentum and energy to be injected into the ISM. In Fig. \ref{fig:OA} we show the final column density map of the central area of Cloud B for both wind models. We also observe bubble structures with the \citet{Agertz2013} prescription, even though the strength of the winds is smaller than in the \citet{Dale2008} case. As expected, the WBB are larger in \citet{Dale2008} as the momentum injection is larger. There is a considerable difference in the number of sinks formed, whereas in \citet{Dale2008} we formed 31 with an average mass of roughly $400 M_{\odot}$, for the \citet{Agertz2013} model we form 208 sinks with an average mass of $167 M_{\odot}$. There are two main morphological differences between the two models. The well defined central cavity created by the strong wind is substituted by a structure of filaments, as the winds do not have the strength to clear up the dense material of this region of the cloud.  The very dense region in the southern part of the figure, is also more affected by the powerful winds of \citet{Dale2008} prescription, and appears less disrupted in the \citet{Agertz2013} model. These effects help to explain the difference in the number of sinks between the two simulations.
 \begin{figure}
 \centering
 \includegraphics[angle=0,width=90mm]{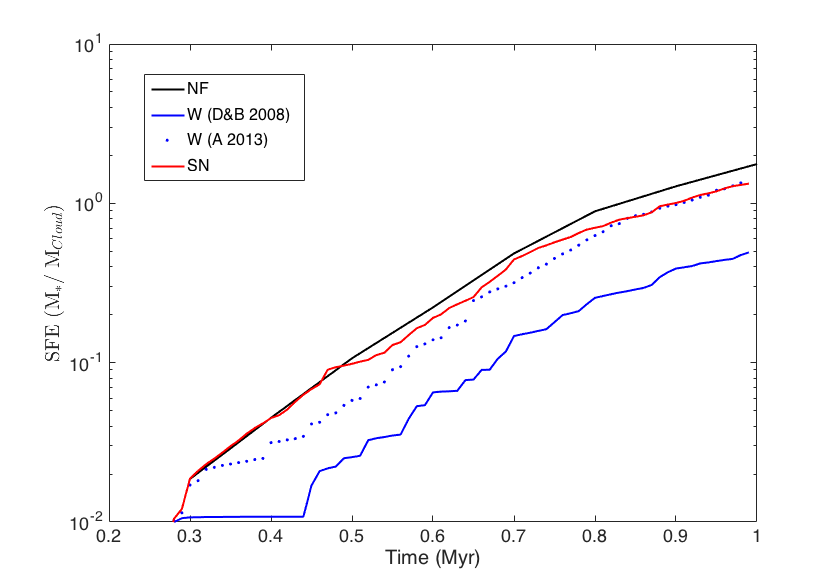}
 \caption{Star Formation Efficiency for Cloud B, comparing the different feedback cases with the non feedback run.}
 \label{fig:SFEOA}
 \end{figure}

We present the SFE for all the simulations of Cloud B in Fig. \ref{fig:SFEOA}. The recipe for winds from \citep{Agertz2013} reduces the SFE, although the reduction is not as pronounced as with the \citet{Dale2008} model. The SFE of the \citet{Agertz2013} recipe is similar to the model with only SNe. However we note here that we are comparing the effect of winds calculated using a realistic population of O/B stars, with an enhanced production of SNe (at least by a factor of 2). 

The increase in the star formation at late times happens in the very dense region on the south of the cloud that is not disrupted by the weak winds in \citet{Agertz2013}. There is a star formation burst in the last 0.2 Myr of the simulation, and the effect of the winds emitted by those new sinks is limited by the simulation time, that ends well before the end of the expected life of the sinks.

\subsection{Tests of the SNe Injection Scheme.}
\begin{figure} 
\centering
\includegraphics[width=80mm]{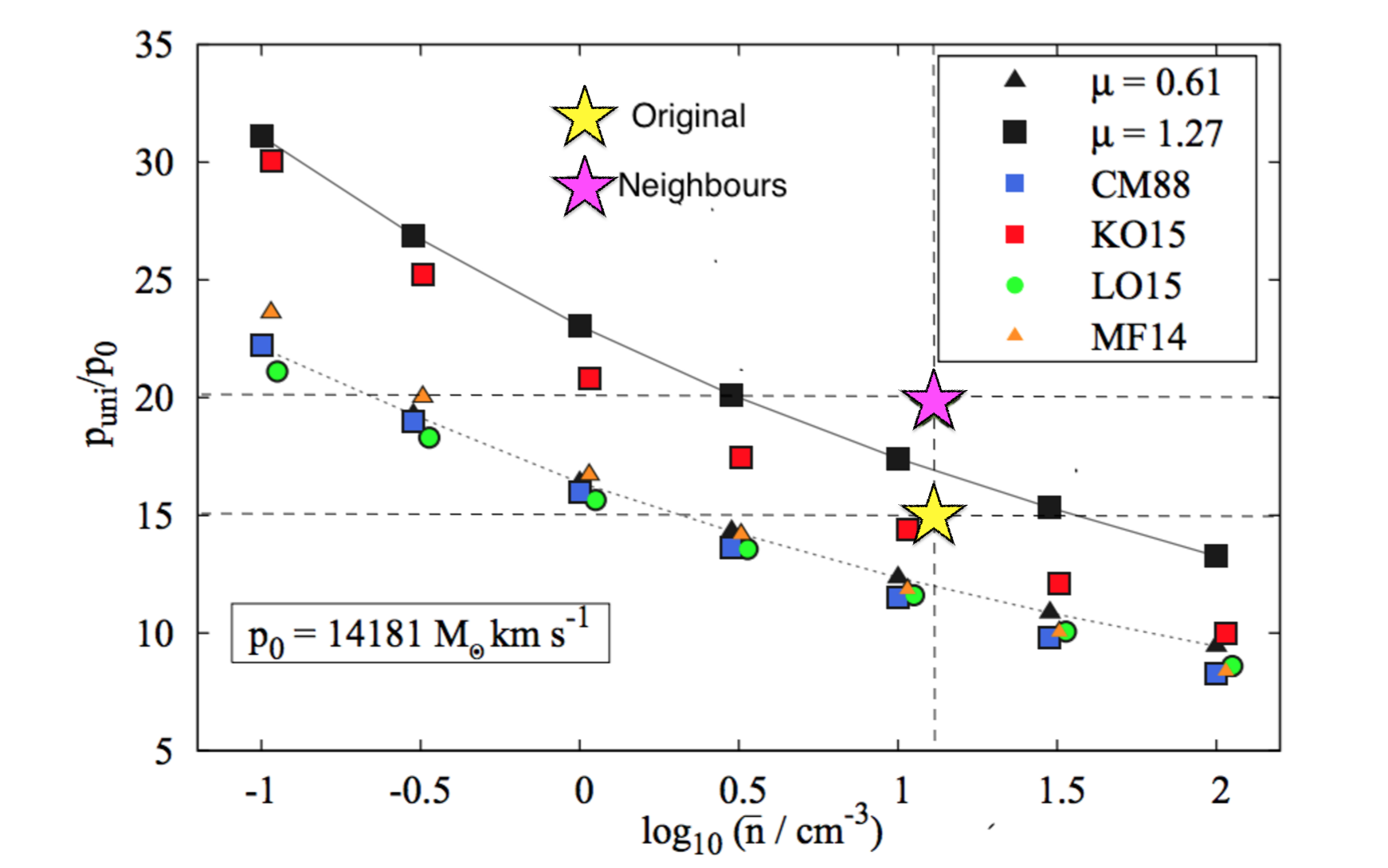}
\caption{Momentum deposition by SNe in a homogeneous medium comparing different methods \citep[adapted from ][]{Haid2016}. For the average densities of our clouds, both methods give a comparable momentum injection.} 
\label{fig:Haid}
\end{figure}

As well as comparing our winds prescription, we also compared  the momentum injection of SNe in our models with \citet{Haid2016}. We used as initial conditions the turbulent sphere with uniform density (from Paper 1) and we set a sink in the centre. Then we test 2 different methods for the SN. In the first method we inject 50\% of $10^{51}$ erg to the 50 neighbours of the sink 100\% as thermal energy. For the second,  we use the method described in the appendix (with a SN$_{\rm eff}$ of 50\%). We calculate the global momentum injection from both and compare it with other models \citep{Haid2016}. The results are shown in the Fig. \ref{fig:Haid} \citep[extracted from][]{Haid2016}, where the yellow star represents our original method, and the purple the neighbour injection. The original method gives better results for the momentum injection lying very close to the results of \citet{KimOstriker2015} (KO15), and 30\% larger than \citet{Cioffi1988} (CM88). We also test the effect of different densities by using a sphere of $10^6$ particles with an average density of n $\sim$ 100 cm$^{-3}$. Although the cooling is very effective and the thermal energy radiates away quickly, we can still observe the formation of a bubble in the centre of the sphere. This is not observed in the neighbours method, where the thermal energy is swiftly radiated away and little effect is observed. In the original method 50\% of the total energy is injected as a momentum \textit{kick}, representing a more suitable way of modelling energy injection in high density environments were cooling is very effective.

\section{Discussion}
There is still considerable uncertainty on the relevance of different feedback processes in molecular clouds. We find that SNe tend to be less effective compared to winds, and the energy and momentum injected by the SNe do not seem to couple to the gas (i.e. dense gas is not so effected by the SNe). This is in agreement with \citet{Rogers2013a}, who claim that 99\% of the energy from the SNe escapes through holes and cavities created by winds. 

Our results are also consistent with \cite{Fierlinger2015} who find SNe are not so important on smaller scales ($\sim$ 10 pc) where winds inject twice as much energy as SNe, whilst \cite{Ibanez-Mejia2015} find that external SNe do not inject enough energy to maintain turbulence. However other studies suggest that SN are the most effective feedback mechanism (\citet{Hensler2010}), and other numerical works claim that SN and photodissociation are more effective than winds \citep{Geen2015,Padoan2015}. One reason SNe may be less effective is that although they heat the gas up to over $10^6$ K, efficient cooling leads to a fairly rapid decrease in temperature. Hot gas emits energetic photons that are able to escape the cloud, using the corridors created in the clouds by the winds \citep{Rogers2013a}, and exporting the energy from the SNe outside the cloud onto the galactic scales. The reduction of the energy coupling between SNe and the gas in the cloud is also observed in other simulations (e.g. \citet{Walch2015}).

We also modelled winds using the \citet{Agertz2013} recipe, which are based on a simple population of stars created at the same time within the sink particle with a fully sampled IMF. This recipe should provide a reasonable estimate of the total momentum input from winds on the cloud. However, we should also note that converting the thermal energy provided by the winds to kinetic energy may underestimate (for large cooling times) or overestimate (for dense regions with short cooling times) the final momentum. The reduction of the SFE on the simulations with winds using \citet{Agertz2013} is now similar to our SNe model (albeit the number of SNe may be superior compared to the \citet{Agertz2013} prescription), thus in this instance there is not so much difference between the effectiveness of winds and supernovae.

A second point, as mentioned above, is that the energy from SNe tends to escape into low density regions, including cavities produced by stellar winds. Some studies which combine ionising radiation and winds \citep[e.g.][]{Ngoumou2014,Dale2014} argue that the addition of winds do not significantly change the effect of the ionising radiation, but we do not include ionising radiation here.

There are differences between our work and the above studies which may further effect the impact of feedback. So for example \citet{Rogers2013a} only model a small patch of a GMC, and do not include self gravity. In our simulations, we model a whole GMC and the inclusion of self gravity likely leads to denser gas and / or a convergent velocity field which counters the effects of feedback. Thus it is unsurprising that feedback appears particularly effective at dispersing the clouds modelled in \citet{Rogers2013a}. Again our spatial scales are larger, and initial conditions much more complex compared to other studies \citep{Dale2014,Geen2015,Fierlinger2015} and even \citet{Walch2012} who model fractal clouds. Furthermore some of these models simply place a star particle emitting feedback at the centre of the simulation \citep{Geen2015,Fierlinger2015}. Our results suggest that the different morphology of clouds may have an impact on the level, and effectiveness of feedback on the gas dynamics of the clouds. 

As well as not including ionising radiation, there are a number of caveats to our models. Firstly we have not simulated our clouds for as long as we would prefer. Partly this is because our initial conditions already contain dense gas, and due to both the density and velocity field of the gas, particles may have small timesteps. Further, sink particles are produced early in the simulations, and the inclusion of winds naturally leads to very small timesteps and increases the computational overhead caused by parallelization. 

We are also aware that we may be overestimating the number of massive stars in our sinks (calculated via the $W_{\rm eff}$ parameter). A relatively large fraction of our sinks are assumed to constitute massive stars, whereas ideally the fraction of massive stars should reflect the observed stellar IMF \citep{Kroupa2001}.  

We are depositing amounts of energy in agreement with other works, but it is worth bearing in mind that lower $W_{\rm eff}$ (and/or assuming fewer massive stars per sink) reduces the effectiveness of winds in our models. However with our simulations we are able to see that the feedback has different effects on clouds with distinct morphologies, whereas models with lower $W_{\rm eff}$ and $SN_{\rm eff}$ either do not present these so clearly, or would require an infeasible simulation time to show differences.

\section{Summary and Conclusions}
In this paper we have performed simulations of molecular clouds that include feedback from winds and SNe. In comparison to previous works, we model GMCs featuring more realistic morphologies. We find that the clouds' inherited properties from the galaxy have a notable influence on the efficacy of feedback. Our main conclusions are:

\begin{enumerate}
\item The structure and velocity field of Cloud B and C are very dissimilar, leading to differences in the strength and effect of the winds. We find that the intrinsic velocity field of the cloud may have an impact on the SFR as significant as other physical properties such as density, mass or the size of the cloud. The velocity field affects the clouds indirectly by modifying the star formation process and therefore the strength of the feedback, for example whether gravitationally collapsing or undergoing shear.\\ 

\item Winds stop accretion onto sinks in both clouds, create wind blown bubbles or regions of lower density, and increase the virial parameter. These effects result in lower SFRs for the clouds. The continuous action of the winds appear more effective in disrupting the clouds than SNe. Both winds and SNe can also lead to overdensities, and trigger star formation in colliding shells or bubbles.\\

\item Stellar feedback is injected locally, and many of the effects we see are relatively local, for example the input of hot gas, changes in the velocity fields and decreases in density of the surrounding gas. However the effects can also lead to a global change in the virial state of the cloud. Feedback appears to act to prevent clouds becoming increasingly bound, having a greater influence on our bound cloud which through the action of winds becomes unbound. With our unbound example, feedback is less effective, but the cloud is always unbound in any case. Furthermore, we find that the overall molecular gas fraction of both clouds remain unaffected by the feedback throughout the duration of our simulations ($t\sim 2$ Myr).\\

\item When studying the transfer of energy from the stellar feedback to the gas in the cloud, we find that only the winds act as an effective source of kinetic and thermal energy. We observe that SNe increase the temperature of the surrounding gas, but this gas is able to cool down swiftly, causing less impact on the cloud.\\
\end{enumerate}

In future work, we will look to include more realistic creation of stellar mass in the sink particles following certain IMF. We are aware that the results in this paper are restricted to two clouds, so we also plan to extend our analysis to a larger number of clouds extracted from galactic simulations.
\section{Acknowledgements}

We would like to thank the anonymous referee for his/her insightful comments that have helped to improve greatly the quality of the paper. The calculations for this Paper were performed on the supercomputer at Exeter, which is jointly funded by STFC, the Large Facilities Capital Fund of BIS and the University of Exeter. RRR and CLD acknowledge funding from the European Research Council for the FP7 ERC starting grant project LOCALSTAR. OA and RRR would like to acknowledge support from STFC consolidated grant ST/M000990/1. Figs. \ref{fig:NF}, \ref{fig:t_evol}, \ref{fig:2CloudsT} and \ref{fig:vfield} were produced using \textsc{splash} \citep{Price2007}. 

\appendix
\section{Stellar feedback model}
\label{appendixA}
\begin{figure*}
\centering
\includegraphics[angle=0,width=150mm]{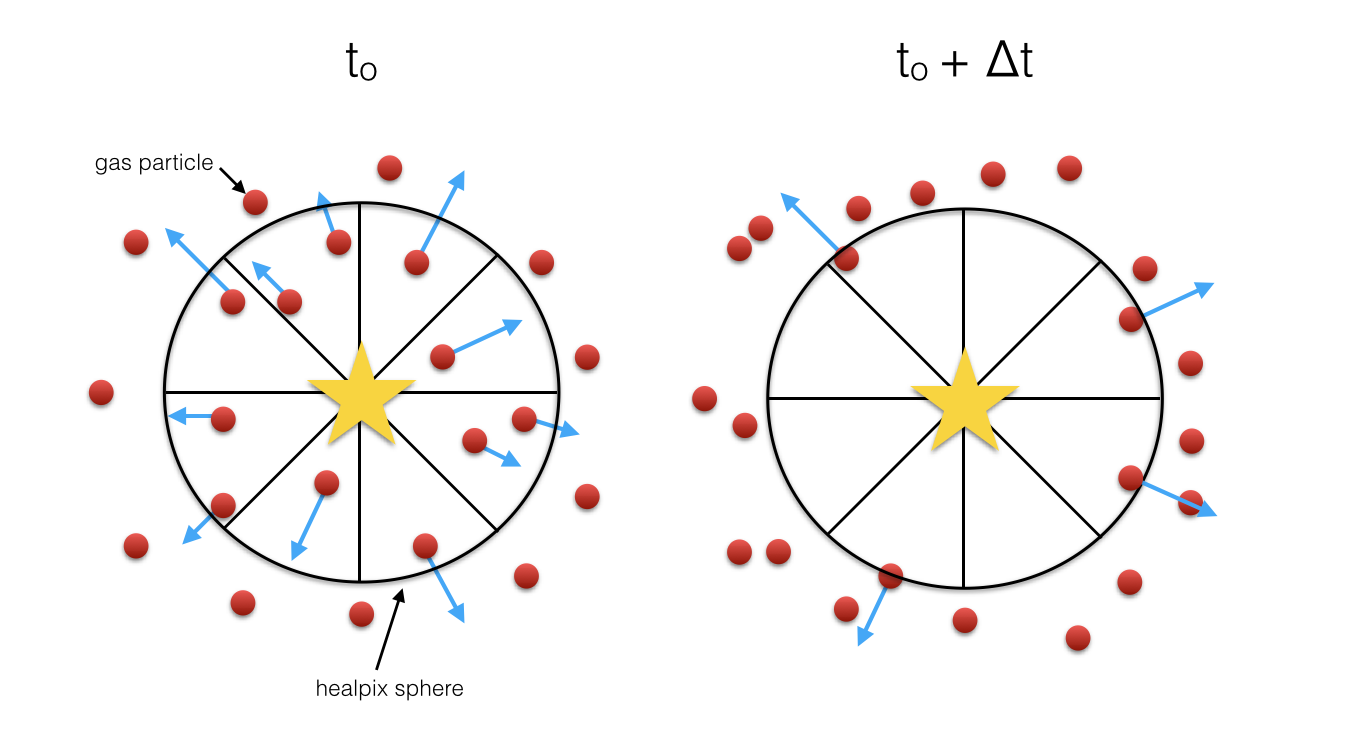}
\caption{Injection of stellar winds using a \textsc{healpix} scheme. In the left panel we portray a distribution of gas particles (represented by red circles) around a sink particle (represented by the star). We show a simplified version of the \textsc{healpix} sphere divided into 8 cells. The total amount of the momentum injection from winds will be divided by the number of pixels or cells. In each cell the corresponding strength of the winds is also divided by the number of particles in the cell. This is represented by large arrows for the cells with only one particle and small arrows for the cells with two particles. The winds point in the outward direction of the sphere. In the right panel, we show the evolution of the system after a small period of time $\Delta$t, some of the gas particles have escaped the sphere and the winds are injected only in the particles that remain within the radius of influence.}
\label{fig:Healpix}
\end{figure*}
To inject feedback we consider a sphere around each sink particle. To account for difference in the gas distribution around sinks and to increase the numerical stability of the code we divide the sphere into different cells using a \textsc{healpix} scheme (\cite{Gorski2005}). \textsc{healpix} divides a sphere in pixels of equal area covering cells of equal volume. The number of pixels scales as $N_{\rm pix} = 12 \cdot N^2_{\rm side}$  where $N_{\rm side}$ is an integer greater or equal to one. We choose $N_{\rm side} = 2$, which yields $N_{\rm pix}= 48$ cells or pixels.\\

\subsection{Injecting Feedback from Winds}
As we cannot resolve individual stars (or an Initial Mass Function), we differentiate between the mass of the sink ($M_{\rm s}$) and the mass of the O/B stars contained in the sink particle ($M_{\rm *}$). To define the stellar mass we include an additional parameter to the simulations $W_{\rm eff}$ fixed at the beginning of the simulation, which represents the fraction of the mass of the sink that will be creating winds
\begin{equation}
M_* = W_{\rm eff}M_{\rm s}.
\label{eq:weff}
\end{equation} 

We only inject feedback if $M_{\rm *}~\ge 8 M_{\odot}$. From the moment a sink particle is created ($t_{\rm *}^{\rm o}$), we inject feedback from winds unless that sink becomes a supernova at $t = t_{\rm *}^{\rm SN}$ ($t_{\rm *}^{\rm SN}$ can again be fixed for a given simulation). To calculate feedback injection from winds we roughly follow \citet{Dale2008}. The winds affect all the particles in a \textsc{healpix} sphere of radius $R_{\rm w}$ around the sink. We performed multiple tests for different values of $R_{\rm w}$, finding that if we choose a small radius, winds would only affect a few particles. These particles will have a huge kick from the feedback which may potentially induce numerical problems. On the other hand if we select a large radius we may be injecting feedback in areas not causally related. For those reasons we consider $R_{\rm w} = 5$ pc (corresponding to the area causally connected on an average timestep). According to \citet{Dale2008} the mass loss rate of an O/B star can be modelled as
\begin{equation}
\label{eq:Dale08}
   \dot{M} = 10^{-5} \left ( \frac{M_*}{30M_{\odot}}\right)^{4} \rm{ M_{\odot} \ yr^{-1}}.
\end{equation}

To compute the momentum increase of each particle in our model ($\Delta \vec{p}_{\rm i}$) we divide the total momentum injection by the number of pixels ($N_{\rm pix}$). Then on each timestep $\Delta t$ we count all the gas particles present ($N_{\rm j}$) in a given cell of the sphere around the sink. 
\begin{equation}
\label{eq:mom}
\Delta p_i = \frac{1}{N_{\rm pix} N_j} \dot{p}_i\ \Delta t,
\end{equation}

For our model, we define $\vec{v}_{\rm w}$ as a parameter representing the terminal velocities of the winds and $\Delta t$ is the current timestep. We set $|\vec{v}_{\rm w}|$  = 1000 km~s$^{-1}$ \citep{Dale2008} and $|p| = \vec{v}_{\rm w} \ \dot{M}$, and choose the direction so that $\vec{v}_{\rm w}$ points radially outwards. The velocity injection of each gas particle in the j-th cell is then 
\begin{equation}
\label{eq:fin_mom}
\Delta \vec{v}_i = \frac{1}{N_{\rm pix} N_j} \vec{v}_w \frac{\dot{M}}{m_p} \ \Delta t,
\end{equation}
$N_{\rm j}$ is the number of particles included in the j-th cell, and $m_{\rm p}$ is the particle mass. We quadratically decrease the amount of injected momentum from the position of the sink to $R_{\rm w}$. This increases the stability of the code, as very distant particles are more likely to have different timesteps. In a typical injection of feedback, we estimate that  80\% of the total energy/momentum is injected in the nearby particles.\\

For the implementation of \citet{Agertz2013} we consider the momentum and energy injection derived from the \textsc{starbust97}. Considering solar metallicity and an age of the star smaller than 6 Myr then the momentum is
\begin{equation}
\label{eq:momOA}
\dot{p}_{\rm w} = 3.6 \cdot 10^{39} \ M_* \ {\rm g \ cm \ s^{-1} \ Myr^{-1}}.
\end{equation}
And the energy is
\begin{equation}
\label{eq:energyOA}
\dot{E}_{\rm w} = 3.8 \cdot 10^{47} \ M_* \ {\rm erg \ Myr^{-1}}.
\end{equation}

We consider that the energy is transformed into kinetic energy and therefore, the total momentum injected by winds is (using Eq. \ref{eq:mom})
\begin{equation}
\label{eq:totpOA}
\dot{p}_i = \dot{p}_{\rm w} + \sqrt{2 m_p  \dot{E}_{\rm w}}.
\end{equation}

\subsection{Modelling Supernovae}
For modelling supernovae we use the same \textsc{healpix} scheme as for the winds. The supernova event takes place only once and when the sink age is larger than $t_{\rm *}^{\rm SN}$. We inject $E_{\rm SN}^{\rm O} = 10^{51}$ erg in a sphere of radius $R_{\rm SN}$. This energy is injected 50\% as momentum deposited on the gas and 50\% as thermal energy. As for the winds, we include $SN_{\rm eff}$ as an extra parameter to vary the efficiency of SNe changing the total energy injected per sink, as\\
\begin{equation}
\label{eq:SN_eff}
E_{SN} = SN_{\rm eff} \cdot E_{SN}^O.
\end{equation}

Following an analysis similar to winds, the momentum injection on each particle is given by
\begin{equation}
\label{eq:SN_mom}
\Delta \vec{v}_i =\sqrt{ \frac{E_{SN}}{N_{\rm pix}~N_{j}~m_{p}}}.
\end{equation}

The thermal energy per unit mass deposited by the supernova event is
\begin{equation}
\label{eq:SN_u}
\Delta e_i =\frac{E_{SN}}{2N_{\rm pix}~N_{j}~m_{p}}.
\end{equation}

A SN event represents a large injection of energy and momentum in one timestep.  We select a larger region for SNe ($R_{\rm SN} = 10$ pc), to reduce numerical effects when injecting a large amount of energy and momentum in just one timestep. We also decrease the amount of injected energy from the position of the sink to $R_{\rm SN}$. Even though there are not many particles with temperatures higher than $T = 10^6$ K, we have set a temperature ceiling of $T_{\rm max} = 10^7$ K in case some particles receive such a large internal energy injection that timesteps are cripplingly small. After undergoing a supernova the sink is no longer emitting feedback.\\ 

\bibliographystyle{mn2e}

\bibliography{Bibliogra}
 
\bsp

\label{lastpage}

\end{document}